\documentclass[12pt,a4paper]{article}
\usepackage{latexsym}

\renewcommand{\title}[1]{\null\vspace{25mm}\noindent{\Large{\bf #1}}\vspace{10mm}}
\newcommand{\authors}[1]{\noindent{\large #1}\vspace{20mm}}
\newcommand{\address}[1]{{\center{\noindent #1\vspace{0mm}}}}
\renewcommand{\abstract}[1]{\vspace{17mm}
\noindent{\small{\em Abstract.} #1}\vspace{2mm}}

\begin{document}

\begin{titlepage}

\begin{center}
\hspace*{\fill}{{\normalsize \begin{tabular}{l}
                              {\sf hep-th/0012112}\\
                              {\sf REF. TUW 00-33.}
                              \end{tabular}   }}

\title{ The Energy-Momentum Tensor on Noncommutative Spaces -
 Some Pedagogical Comments}

\vspace{10mm}

\authors{  \large{A. Gerhold$^{1}$, 
 J. Grimstrup$^{2}$ ,
 H. Grosse$^{3}$, 
 L. Popp$^{4}$ ,\\ M. Schweda$^{5}$, 
 R. Wulkenhaar$^{6}$}}

\vspace{5mm}

\address{$^{1,2,4,5}$  Institut f\"ur Theoretische Physik, Technische Universit\"at Wien\\
      Wiedner Hauptstra\ss e 8-10, A-1040 Wien, Austria}
\address{$^{3,6}$  Institut f\"ur Theoretische Physik, Universit\"at Wien\\Boltzmanngasse 5, A-1090 Wien, Austria   }
\footnotetext[2]{Work supported by The Danish Research Agency.}
\footnotetext[4]{Work supported in part by ``Fonds zur F\"orderung der Wissenschaftlichen Forschung'' (FWF) under contract P13125-PHY and P13126-PHY.}
\footnotetext[6]{Marie-Curie Fellow.}       
\end{center} 
\thispagestyle{empty}
\begin{center}
\begin{minipage}{12cm}

\vspace{10mm}

{\it Abstract.} We present the discussion of the energy-momentum tensor of the scalar $\phi^4$-theory on a noncommutative space. The Noether procedure is performed at the 
operator level. Additionally, the broken dilatation symmetry will be considered
in a Moyal-Weyl deformed scalar field theory at the classical level.

\end{minipage}\end{center}
\end{titlepage}

\section {Introduction}
The aim of this work is to investigate the translation and dilatation symmetry
at least at the classical level for a noncommutative $\phi^4$-theory. Not
much work has yet been done in this direction, only in \cite{r1} and \cite{r2}
one finds some scattered remarks concerning the energy-momentum tensor and its
Noether procedure for Moyal-Weyl deformed scalar field theories. In
this paper we extend the analysis of \cite{r2} and formulate the Noether 
procedure for translations already at the operator level. By the use of the 
Moyal-Weyl correspondence between operators and  fields we are able
to confirm the results of \cite{r2}.

This work is organized as follows. Section 2 is devoted to some special
features of the quantum space in connection with a $\phi^4$-theory.

In Section 3 we study the construction of the energy momentum tensor at the
operator level and in a Moyal-Weyl deformed $\phi^4$-theory.

Finally, in the last section we investigate the broken dilatation symmetry of
the noncommutative $\phi^4$-theory.

\section {The quantum phase space and the scalar field theory}

We consider the scalar field theory which is described at the classical
level by the following action\footnote{We use the shorthand notations 
$dx:=d^4x$ and $dk:={d^4k\over(2\pi)^4}$.}
\begin {equation}
 S^{(0)}[\phi]=\int dx\,\left({1\over2}\partial_\mu\phi\partial_\mu\phi
 +{m^2\over2}\phi^2+{\lambda\over4!}\phi^4\right),
\end{equation}
where $\phi(x)$ is a real valued scalar field on the four dimensional
Euclidean space time $E_4$. For fields in a Schwartz space of functions which
decrease sufficiently fast at infinity we may define a Fourier transformation
by
\begin{eqnarray}
 \phi(x)&=&\int dk\, e^{ik_\mu x_\mu}\tilde{\phi}(k), \nonumber\\
 \tilde{\phi}(k)&=&\int dx\,e^{-ik_\mu x_\mu}\phi(x), \label{l2}
\end{eqnarray}
with $\tilde{\phi}(-k)=\tilde{\phi}^*(k)$. In order to generalize a field 
theory on an ordinary space-time to one on a noncommutative space-time we 
replace the local coordinates $x_\mu$ by hermitian operators $\hat{x}_\mu$
obeying the relations
\begin{equation}
 [\hat{x}_\mu,\hat{x}_\nu]=i\sigma_{\mu\nu},\quad [\hat{x}_\mu,\sigma_{\mu\nu}]
 =0, \label{l3}
\end{equation}
where $\sigma_{\mu\nu}=-\sigma_{\nu\mu}$ is a real invertible matrix.
Consequently, fields
on space-time are replaced by operators. Replacing $x_\mu$ by $\hat{x}_\mu$
in (\ref{l2}) we obtain 
\begin{equation}
 \phi(\hat{x}_\mu)=\int dk\,e^{ik\hat{x}}\tilde{\phi}(k).
\end{equation}
With (\ref{l2}) one gets
\begin{eqnarray}
 \phi(\hat{x})&=&\int dk \int dx\,\phi(x)e^{ik\hat{x}-ik x}=
 \int dx \int dk\,\hat{T}(k)e^{-ik x}\phi(x)=\nonumber\\
 &=&\int dx \hat{\Delta}(x)\phi(x). \label {l5}
\end{eqnarray}
$\phi(\hat{x})$ is an element of an algebra $\mathcal{A}_x$ in the sense of
\cite{r6}.

In (\ref{l5}) we have introduced the operators $\hat{T}(k)$ and $\hat{\Delta}
(k)$ which were originally defined by Balasz et al. \cite{r3}.
More recently, these operators were also used by Filk \cite{r4}:
\begin{equation}
 \hat{T}(k)=e^{ik\hat{x}},
\end{equation}
and by Ambjorn et al. \cite{r5}:
\begin{equation}
 \hat{\Delta}(x)=\int dk\,e^{ik \hat{x}-ik x}=
 \int dk\,\hat{T}(k)e^{-ik x}.
\end{equation}
$\hat{T}(k)$ and $\hat{\Delta}(x)$ have different useful properties for 
practical
calculations. In order to list these properties let us define the trace
operations for $\hat{T}(k)$ and $\hat{\Delta}(x)$. 

For simplicity we choose the space time dimension $d=2$ and consider in a first
step the trace of $\hat{T}(k)$. The operator $\hat{T}(k)$ has the following 
properties
\cite{r4}:
\begin{eqnarray}
 \hat{T}^\dag(k)&=&\hat{T}(-k) \nonumber\\
 \hat{T}(k)\hat{T}(k^\prime)&=&e^{-ik\times k^\prime}\hat{T}(k+k^\prime),
\end{eqnarray}
where $k\times k^\prime := {1\over2}\sigma_{\mu\nu}k_\mu k_\nu^\prime.$
For $d=2$ we have
\begin{equation}
 \sigma_{\mu\nu}=\sigma \varepsilon_{\mu\nu}=\sigma\left( \begin{array}{cc}
 0&1\\-1&0\end{array}\right),
\end{equation}
and eq. (\ref{l3}) becomes
\begin{equation}
 [\hat{x}_1,\hat{x}_2]=i\sigma. \label{l10}
\end{equation}
The following remarks concerning the definitions of traces can be deduced with 
the methods of \cite{r3}. Eq. (\ref{l10}) looks like the usual commutation
relation of ordinary quantum mechanics between $\hat{q}$ and $\hat{p}$ if one
identifies $\hat{x}_1=\hat{q}$ and $\hat{x}_2=\hat{p}$. The corresponding 
eigenstates are defined by 
\cite{r3}:
\begin{eqnarray}
 \hat{x}_1|x\rangle&=&x|x\rangle, \nonumber\\
 \hat{x}_2|p\rangle&=&\sigma p|p\rangle
\end{eqnarray}
with
\begin{eqnarray}
 &&\langle x|x^\prime\rangle=\delta(x-x^\prime),\quad \int dx\,|x\rangle\langle
 x|=1,\nonumber\\
 &&\langle p|p^\prime\rangle=\delta(p-p^\prime),\quad \int dp\,|p\rangle\langle
 p|=1,
\end{eqnarray}
and
\begin{equation}
 \langle x|p\rangle={1\over\sqrt{2\pi}}e^{ipx}.
\end{equation}
Now it is straightforward to calculate the matrix elements of $\hat{T}(k)$.
The result is
\begin{equation}
 \langle x^\prime|\hat{T}(k)|x^{\prime\prime}\rangle=\delta(k_2\sigma+x^\prime
 -x^{\prime\prime})e^{ik_1(x^\prime+x^{\prime\prime})/2}, \label{l13}
\end{equation}
implying that the trace is (with an appropriate normalization factor)
\begin{eqnarray}
 \mathrm{Tr}\,\hat{T}(k)&:=&2\pi\sigma\int dx\,\langle x|T(k)|x\rangle= 
 \nonumber\\
 &=&2\pi\sigma\,\delta(k_2\sigma)\int dx\,e^{ik_1x}= (2\pi)^2\delta^{(2)}
 (k_\mu).\label{l14}
\end{eqnarray} 
With eq. (\ref{l13}) we are also able to calculate the matrix elements of
$\hat{\Delta}(x)$. A short calculation gives
\begin{eqnarray}
 \langle x^\prime|\hat{\Delta}(x)|x^{\prime\prime}\rangle&=&
 \int dk\,\langle x^\prime|\hat{T}(k)|x^{\prime\prime}\rangle e^{-ik x}=
 \nonumber\\
 &=&{1\over2\pi\sigma}\,\delta\left(x_1-{x^\prime+x^{\prime\prime}\over2}
 \right)e^{i(x^\prime-x^{\prime\prime})x_2/\sigma}
\end{eqnarray}
and the trace of $\hat{\Delta}(x)$ becomes
\begin{equation}
 \mathrm{Tr}\,\hat{\Delta}(x):=2\pi\sigma\int dx\, \langle x|
 \hat{\Delta}(x)|x\rangle=\int dx\,\delta(x_1-x)=1. \label{l16}
\end{equation}
Eqs.(\ref{l14}) and (\ref{l16}) confirm the results of \cite{r4,r5}.
Additionally, one can derive the following relations
\begin{eqnarray}
 \mathrm{Tr}\,[\hat{T}(k)\hat{T}(k^\prime)]&=&(2\pi)^2e^{-ik\times k^\prime}
 \delta^{(2)}(k_\mu+k^\prime_\mu)=(2\pi)^2\, \delta^{(2)}(k_\mu+k^\prime_\mu),
 \nonumber\\
 \mathrm{Tr}\,[\hat{\Delta}(x)\hat{\Delta}(x^\prime)]&=&
 \delta^{(2)}(x_\mu-x^\prime_\mu).
\end{eqnarray}
In order to be complete, we present an alternative way of calculating the
trace of $\hat{T}(k)$:
\begin{equation}
 \mathrm{Tr}\,\hat{T}(k)=2\pi\sigma\int dx^\prime\,\langle x^\prime|\hat{T}(k)
 |x^\prime\rangle,
\end{equation}
where $|x^\prime\rangle$ is now an appropriate representation of the algebra
(\ref{l3}):
\begin{equation}
 [\hat{x}_\mu,\hat{x}_\nu]|x^\prime\rangle=i\sigma\varepsilon_{\mu\nu}|x^\prime
 \rangle. \label{l19}
\end{equation}
A possible solution for (\ref{l19}) is
\begin{equation}
 \hat{x}_\mu|x^\prime\rangle=\left(x_\mu+{i\over2}\sigma_{\mu\rho}\partial_\rho
\right)|x^\prime\rangle.
\end{equation}
However, in this case $x^\prime$ cannot be identified with $x_\mu$ due to the
fact that $x_\mu$ and $\partial_\rho$ represent $2\times d$ degrees of freedom
($d=2$). Therefore we need an irreducible representation which eliminates the
redundant degrees of freedom.

An irreducible representation is given by (renaming $x^\prime\to x$):
\begin{eqnarray}
 \hat{x}_1|x\rangle&=&x|x\rangle \nonumber\\
 \hat{x}_2|x\rangle&=&-i\sigma{d\over dx}|x\rangle.
\end{eqnarray}
Using the Baker-Campbell-Hausdorff-formula and the fact that
\begin{equation}
 e^{\sigma k_2{d\over dx}}|x\rangle=|x-\sigma k_2\rangle,
\end{equation}
one obtains again the result (\ref{l14}).

In order to define a scalar field theory at the operator level we need a 
derivation prescription \cite{r5,r6,r7}:
\begin{equation}
 \hat{\partial}_\mu\phi(\hat{x})=
 -i[\hat{x}^\prime_\mu,\phi(\hat{x})]=\int dx\,\partial_\mu\phi(x)
 \hat{\Delta}(x), \label{l30}
\end{equation}
where $\hat{x}^\prime_\mu=\sigma^{-1}_{\mu\nu}\hat{x}_\nu$ and 
$\sigma_{\mu\rho}\sigma^{-1}_{\rho\nu}=\delta_{\mu\nu}$.
This definition implies
\begin{equation}
 [\hat{\partial}_\mu,\hat{x}_\nu]=\delta_{\mu\nu},\quad [\hat{\partial}_\mu,
 \hat{\partial}_\nu]=0.
\end{equation}
Furthermore, we have the Leibniz rule
\begin{equation}
 \hat{\partial}_\mu(f(\hat{x})g(\hat{x}))=\hat{\partial}_\mu f(\hat{x})
 g(\hat{x})+f(\hat{x})\hat{\partial}_\mu g(\hat{x}).
\end{equation}
Additionally, one can show that one has the following useful relation
\begin{equation}
% [\hat{\partial}_\mu,\phi(\hat{x})]=\int dx\,\partial_\mu\phi(x)\hat{\Delta}
% (x), \label{l26}\\
 \big[\hat{\partial}_\mu,\hat{\Delta}(x)\big]=-\partial_\mu\hat{\Delta}(x).
 \label{l31}
\end{equation}
Eq. (\ref{l31}) implies
\begin{equation}
 e^{-v_\mu\hat{\partial}_\mu}\hat{\Delta}(x)e^{v_\mu\hat{\partial}_\mu}
 =\hat{\Delta}(x+v).\label{l32}
\end{equation}
The existence of such an operator implies that $\mathrm{Tr}\,\hat{\Delta}(x)$
is independent of $x$ for any trace operation on the algebra of operators.
(\ref{l32}) gives therefore
\begin{equation}
 \mathrm{Tr}\,\hat{\Delta}(x)=\mathrm{Tr}\hat{\Delta}(x+v)
\end{equation}
and thus one has in consistency with (\ref{l16}):
\begin{equation}
 \mathrm{Tr}\,\phi(\hat{x})=\int dx\,\phi(x)\mathrm{Tr}\,\hat{\Delta}(x)=
 \mathrm{Tr}\,\hat{\Delta}(x)\int dx\,\phi(x).
\end{equation}
In normalizing $\mathrm{Tr}\,\hat{\Delta}(x)$ to one we get
\begin{equation}
 \mathrm{Tr}\,\phi(\hat{x})=\int dx\,\phi(x).
\end{equation}
Now we are able to define the inverse map of (\ref{l5}). In Filk's \cite{r4}
notation one obtains
\begin{equation}
 \phi(x)=\int dk\,e^{ik x}\,\mathrm{Tr}\,[\phi(\hat{x})T^\dag(k)]
\end{equation}
and corresponding to Ambjorn et al. \cite{r5} one has
\begin{equation}
 \phi(x)=\mathrm{Tr}[\phi(\hat{x})\hat{\Delta}(x)], \label{l37}
\end{equation}
allowing now to define a Moyal-Weyl product \cite{r4,r5} in the following
manner
\begin{eqnarray}
 (\phi_1*\phi_2)(x)&:=&\int dk\,e^{ik x}\,\mathrm{Tr}\,[\phi_1(\hat{x})
 \phi_2(\hat{x})T^\dag(k)]=\mathrm{Tr}\,[\phi_1(\hat{x})\phi_2(\hat{x})
 \hat{\Delta}(x)] \nonumber\\
 &=&\int dk_1\int dk_2\,e^{i(k_1+k_2) x}e^{-ik_1\times k_2}\tilde{\phi}_1
 (k_1)\tilde{\phi}_2(k_2). \label{l38}
\end{eqnarray}
Eqs. (\ref{l37}) and (\ref{l38}) show that there is a one-to-one correspondence
between fields (of sufficiently rapid decrease at infinity) and operators.
From (\ref{l38}) follows also
\begin{equation}
 \int dx\,(\phi_1*\phi_2)(x)=\int dx\,\phi_1(x)\phi_2(x).
\end{equation}
Furthermore one has
\begin{equation}
 \mathrm{Tr}\,[\phi_1(\hat{x})\phi_2(\hat{x})]=\int dx\,(\phi_1*\phi_2)(x).
\end{equation}
and
\begin{equation}
 \mathrm{Tr}[\phi(\hat{x})^4]=\int dx\,(\phi(x))^4_*.
\end{equation}
One can easily show that cyclic rotation is allowed:
\begin{equation}
 \int dx\,(\phi_1*\phi_2*\ldots*\phi_n)(x)=
 \int dx\,(\phi_n*\phi_1*\ldots*\phi_{n-1})(x).
\end{equation}
Using now all these definitions one is able to define a scalar field theory
on a noncommutative space-time at the ``algebra'' level as
\begin{eqnarray}
 \bar{S}^{(0)}[\phi]%&=&\mathrm{Tr}\left(-{1\over2}[\hat{x}_\mu^\prime,
 %\phi(\hat{x})]^2+{m^2\over2}\phi(\hat{x})^2+{\lambda\over4!}\phi(\hat{x})^4
 %\right)=\nonumber\\
 &=&\mathrm{Tr}\left({1\over2}(\hat{\partial}_\mu
 \phi(\hat{x}))^2+{m^2\over2}\phi(\hat{x})^2+{\lambda\over4!}\phi(\hat{x})^4
 \right)=\nonumber\\
 &=&\mathrm{Tr}\,\left(\bar{\mathcal{L}}^{(0)}(\phi(\hat{x}))\right).
\end{eqnarray}
With help of (\ref{l30}) the latter expression may be 
rewritten as a ``Moyal-Weyl deformed'' action:
\begin{eqnarray}
 S^{(0)}[\phi]&=&\int dx\left({1\over2}\partial_\mu\phi*\partial_\mu\phi+
 {m^2\over2}\phi*\phi+{\lambda\over4!}(\phi)_*^4\right)=\nonumber\\
 &=&\int dx\left({1\over2}\partial_\mu\phi\partial_\mu\phi+
 {m^2\over2}\phi^2+{\lambda\over4!}(\phi)_*^4\right)=\nonumber\\
 &=&\int dx\,\mathcal{L}^{(0)}_*(\phi(x)). \label{l245}
\end{eqnarray}
We conclude this section with some remarks concerning the equation of motion
at the algebra level. In order to see how this works it is sufficient to 
discuss the free kinetic part
\begin{equation}
 \bar{S}^{(0)}_{free}[\phi]=-{1\over2}\mathrm{Tr}\left([\hat{x}_\mu^\prime,
 \phi(\hat{x})][\hat{x}^\prime_\mu,\phi(\hat{x})]\right)
 ={1\over2}\mathrm{Tr}\,(\hat{\partial}_\mu\phi(\hat{x}))^2.
\end{equation}
The ``classical'' equation of motion, similar to the commutative case,
is obtained by minimizing the action:
\begin{equation}
 {\delta \bar{S}^{(0)}_{free}[\phi] \over \delta\phi(\hat{x})}=0.
\end{equation}
We define the functional derivative as usual \cite{r2}:
\begin{equation}
 \bar{S}^{(0)}_{free}[\phi+\delta\phi]-
 \bar{S}^{(0)}_{free}[\phi]=:
 \mathrm{Tr}\left({\delta \bar{S}^{(0)}_{free}[\phi] \over \delta\phi
 (\hat{x})} \delta\phi(\hat{x})\right).
\end{equation}
Using cyclic rotation we obtain
\begin{equation}
 {\delta \bar{S}^{(0)}_{free}[\phi] \over \delta\phi(\hat{x})}=
 [\hat{x}_\rho^\prime,[\hat{x}_\rho^\prime,\phi(\hat{x})]]=
 -\hat{\partial}_\rho\hat{\partial}_\rho\phi(\hat{x})=0.
\end{equation}
This is the massless free field equation of the theory. The inclusion of
the mass term
and the interaction gives the following equation of motion:
\begin{equation}
 {\delta \bar{S}^{(0)}[\phi] \over \delta\phi(\hat{x})}=
 -\hat{\partial}_\rho\hat{\partial}_\rho\phi(\hat{x})+m^2\phi(\hat{x})+
 {\lambda\over3!}\phi(\hat{x})^3=0. \label{l46}
\end{equation}
Eq. (\ref{l46}) will be used for the construction of the energy momentum
tensor in the next section. For the Moyal-Weyl deformed field theory one gets
in a similar way the equation of motion \cite{r2}
\begin{equation}
 {\delta S^{(0)}[\phi(x)] \over \delta\phi(x)}=
 -\partial_\rho\partial_\rho\phi(x)+m^2\phi(x)+
 {\lambda\over3!}(\phi)_*^3(x)=0.
\end{equation}

\section{Noether theorem for translation symmetry at the algebra level
and its Moyal-deformed counterpart}
In order to define infinitesimal translations at the operator level one 
generalizes the usual transformation law for a scalar field
\begin{equation}
 \delta_\mu\phi(x)=\partial_\mu\phi(x)%,\quad W_\mu=\int dx\,\partial_\mu\phi
\end{equation}
into
\begin{equation}
 \delta_\mu\phi(\hat{x})=\hat{\partial}_\mu\phi(\hat{x})=-i[\hat{x}_\mu^\prime,
 \phi(\hat{x})]
\end{equation}
in accordance with (\ref{l30}). Since the action 
\begin{equation}
 \bar{S}^{(0)}[\phi]=\mathrm{Tr}\left({1\over2}(\hat{\partial}_\mu
 \phi(\hat{x}))^2+{m^2\over2}\phi(\hat{x})^2+{\lambda\over4!}\phi(\hat{x})^4
 \right)
\end{equation}
is invariant under translations we can try to derive a Noether current in the
following way. One calculates the variation of $\bar{S}^{(0)}[\phi(\hat{x})]$ 
in two different ways, once using the equation of motion and alternatively
without using the equation of motion \cite{rj}.

First we note that with help of (\ref{l30}) and performing cyclic 
rotations under the trace one obtains the following
formula for ``partial integration''
\begin{equation}
 \mathrm{Tr}\left(\phi_1(\hat{x})\hat{\partial}_\mu\phi_2(\hat{x})\right)=
 -\mathrm{Tr}\left(\hat{\partial}_\mu\phi_1(\hat{x})\phi_2(\hat{x})\right).
\end{equation}
\pagebreak
Then we have
\begin{eqnarray}
 &&\delta_\mu \bar{S}^{(0)}[\phi]_1=\nonumber\\
 &&\quad=\mathrm{Tr}\Big(\hat{\partial}_\mu\hat{\partial}_\rho\phi(\hat{x})
 \hat{\partial}_\rho
 \phi(\hat{x})+m^2\hat{\partial}_\mu\phi(\hat{x})\phi(\hat{x})+{\lambda\over3!}
 \hat{\partial}_\mu\phi(\hat{x})\phi^3(\hat{x})\Big)=\nonumber\\
 &&\quad=\mathrm{Tr}\,\Big(\hat{\partial}_\mu\mathcal{L}(\phi(\hat{x}))\Big),
 \nonumber\\\
 &&\delta_\mu \bar{S}^{(0)}[\phi]_2=\nonumber\\
 &&\quad\mathrm{Tr}\Big(\hat{\partial}_\rho(\hat{\partial}_\mu\phi(\hat{x})
 \hat{\partial}_\rho
 \phi(\hat{x}))
 +\hat{\partial}_\mu\phi(\hat{x})\Big(-\hat{\partial}_\rho
 \hat{\partial}_\rho\phi
 (\hat{x})+m^2\phi(\hat{x})+{\lambda\over3!}\phi(\hat{x})^3\Big)\Big).
 \nonumber\\
\end{eqnarray}
Clearly, one has for the difference
\begin{equation}
 \delta_\mu \bar{S}^{(0)}[\phi]_1-\delta_\mu \bar{S}^{(0)}[\phi]_2=0.
\end{equation}
This leads to
\begin{equation}
 \mathrm{Tr}\,\Big(\hat{\partial}_\rho\Big[{1\over2}\big(\hat{\partial}_\rho
 \phi(\hat{x})
 \hat{\partial}_\mu\phi(\hat{x})+\hat{\partial}_\mu\phi(\hat{x})
 \hat{\partial}_\rho\phi(\hat{x})\big)-\delta_{\rho\mu}
 \bar{\mathcal{L}}(\phi(\hat{x}))\Big]\Big)
 =\mathrm{Tr}\,\Big(\hat{\partial}_\rho T_{\rho\mu}\Big)=0. \label{l310}
\end{equation}
where we have defined the (symmetrized) energy-momentum tensor at the
algebra level
\begin{equation}
 T_{\rho\mu}(\phi(\hat{x})):= {1\over2}\big(\hat{\partial}_\rho\phi(\hat{x})
 \hat{\partial}_\mu\phi(\hat{x})+\hat{\partial}_\mu\phi(\hat{x})
 \hat{\partial}_\rho\phi(\hat{x})\big)-\delta_{\rho\mu}\bar{\mathcal{L}}^{(0)}.
 \label{l311}
\end{equation}
It is important to note that eq. (\ref{l310}) does not imply
$\hat{\partial}_\rho T_{\rho\mu}=0$ locally.

For the further discussion we switch to Minkowski space $M_4$.
Using the Moyal-Weyl prescription one can rewrite (\ref{l311}) as\footnote{ In \cite{r12} one finds some further useful remarks concerning translation symmetry in deformed quantum field theories.}                             
\begin{equation}
 T_{\rho\mu}(\phi(x))= {1\over2}\big(\partial_\rho\phi*
 \partial_\mu\phi+\partial_\mu\phi*
 \partial_\rho\phi\big)-\eta_{\rho\mu}\mathcal{L}_*^{(0)}. \label{l57}
\end{equation}
The construction (\ref{l57}) is symmetric - therefore no
Belinfante-procedure is needed \cite{r8}. 
The result (\ref{l57}) is consistent with
\begin{equation}
 W_\mu S^{(0)}[\phi]=\int dx\,\partial_\mu\phi*
 {\delta S^{(0)}[\phi]\over\delta\phi(x)}
 =\int dx\,\partial^\rho T_{\rho\mu}=0. \label{l60}
\end{equation}
We add an improvement 
term in order to get an improved energy-momentum tensor which is 
traceless for $m=0$ \cite{r8}:
\begin{equation}
 T^I_{\rho\mu}=T_{\rho\mu}+{1\over6}(\eta_{\rho\mu}\Box-\partial_\rho
 \partial_\mu)(\phi*\phi). \label{l61}
\end{equation}
The improvement term does not contribute to the divergence of the 
energy-momentum tensor which is given by
\begin{equation}
 \partial^\rho T_{\rho\mu}= \partial^\rho T^I_{\rho\mu}=
 {\lambda\over4!}[[\phi,\partial_\mu\phi]_\textrm{\tiny{\emph{M}}},
 \phi*\phi]_\textrm{\tiny{\emph{M}}} \neq0, \label{l59}
\end{equation}
where we have introduced the Moyal bracket
\begin{equation}
 [\phi_1(x),\phi_2(x)]_\textrm{\tiny{\emph{M}}}:=(\phi_1*\phi_2)(x)-
 (\phi_2*\phi_1)(x).
\end{equation}
The result (\ref{l59}) is already given in \cite{r2}

For a physical interpretation one chooses $\sigma^{0i}=0$ \cite{r1,r2}. Then
one has\footnote{Note that the $\sigma_{\mu\nu}$-matrix is no longer 
invertible, and therefore we restrict our attention to the Moyal-deformed
field theory.}
\begin{equation}
 \int d^3x\,(\phi_1*\phi_2*\ldots*\phi_n)(x)=
 \int d^3x\,(\phi_n*\phi_1*\ldots*\phi_{n-1})(x).
\end{equation}
Eq. (\ref{l59}) implies 
\begin{eqnarray}
 &&\int d^3x\,\partial^\rho T_{\rho\mu}=
 \partial^0\int d^3x\,T_{0\mu}+\int d^3x \partial^i T_{i\mu}=
 \partial^0\int d^3x\,T_{0\mu}= \nonumber\\
 &&=\int d^3x\,{\lambda\over4!}[[\phi,\partial_\mu\phi]_
 \textrm{\tiny{\emph{M}}},\phi*\phi]_\textrm{\tiny{\emph{M}}}=0
\end{eqnarray}
which means that in this case there exists a conserved four momentum:
\begin{equation}
 \partial^0P_\mu:=\partial^0\int d^3x\,T_{0\mu}=0.
\end{equation}
Additionally, $\sigma^{0i}=0$ allows to establish unitarity \cite{r9}.

As it is well known, in the commutative case  the generators of the 
conformal group are given by moments of the energy-momentum tensor \cite{r8}.
E.g. in the commutative case the conserved current for dilatation symmetry is 
given by
\begin{equation}
 D_\mu=x^\rho T^I_{\rho\mu}.
\end{equation}
However, in the noncommutative case one expects a breaking of the dilatation 
symmetry due to the fact that the energy-momentum tensor is not conserved.
As a simple example we study in the last section the broken dilatation 
symmetry in a Moyal-Weyl deformed field theory.

\section{The broken dilatation symmetry}
In this section we express the dilatation transformation in terms of 
a functional differential operator, i.e. we consider
\begin{equation}
 W_D=\int dx\,\delta_D\phi*{\delta\over\delta\phi(x)}=
 \int dx\,(1+x^\mu*\partial_\mu)\phi*{\delta\over\delta\phi(x)}
\end{equation}
acting on the Minkowskian action $S^{(0)}[\phi]$ for a massless field given by
\begin{equation}
 S^{(0)}=\int dx\,\left({1\over2}\partial_\rho\phi\partial^\rho\phi-
 {\lambda\over4!}(\phi)_*^4\right).
\end{equation}
Using
\begin{eqnarray}
 x^\mu&=&(2\pi)^4\int dp\,e^{ip x}\,i{\partial\over\partial p_\mu}
 \delta^{(4)}(p), \nonumber\\
 \partial_\mu\phi(x)&=&\int dp\,e^{ip x}\,ip_\mu\tilde{\phi}(p)
\end{eqnarray}
one verifies with the definition of the Moyal product (\ref{l38})
\begin{equation}
 x^\mu*\partial_\mu\phi(x)=x^\mu\partial_\mu\phi(x).
\end{equation}
%With a short calculation one can prove the following identity:
%\begin{equation}
% \int dy\, \partial_\mu\phi(y)*{\delta\over\delta\phi(y)}(\phi)_*^4(x)=
% W_\mu(\phi)_*^4=\partial_\mu(\phi)_*^4
%\end{equation}
%where $W_\mu$ is given by eq. (\ref{l60}).
Then one gets using the
improved energy-momentum tensor (\ref{l61})
\begin{eqnarray}
 &&W_D S^{(0)}[\phi]=-\int dx\bigg[ \partial^\rho\left(x^\mu*T^I_{\rho\mu}
 \right)+\nonumber\\
 &&\quad +{1\over2}(\phi*\Box\phi-\Box\phi*\phi)+{1\over2}
 x^\mu*(\partial_\mu\phi*\Box\phi-\Box\phi*\partial_\mu\phi)+ \nonumber\\
 &&\quad +{\lambda\over4!} x^\mu*\left(4\partial_\mu\phi*(\phi)_*^3-
 \partial_\mu(\phi)_*^4\right)\bigg]. \label{l73}
\end{eqnarray}
It is straightforward to show that the terms in the second line of (\ref{l73})
vanish and thus one has
\begin{equation}
 W_D S^{(0)}[\phi]=-\int dx\bigg[ \partial^\rho\left(x^\mu*T^I_{\rho\mu}
 \right)+\underbrace{{\lambda\over4!} x^\mu*\left(4\partial_\mu\phi*(\phi)_*^3-
 \partial_\mu(\phi)_*^4\right)}_{=:B}\bigg]. \label{l75}
\end{equation}
%In the commutative case ($\sigma^{\mu\nu}=0$) the breaking $B$ vanishes.
A rather lenghty but straightforward calculation shows that the breaking
$B$ can be written as
\begin{equation}
 B=-2\sigma^{\mu\nu}{\partial S^{(0)}[\phi]\over\partial\sigma^{\mu\nu}}
 \label{l76}
\end{equation}
which demonstrates that the breaking is determined by the deformation parameter
$\sigma^{\mu\nu}$.
The result (\ref{l76}) can be understood in the following way. An 
``infinitesimal'' dilatation
\begin{equation}
 \hat{x}^{\prime\mu}=(1+\varepsilon)\hat{x}^\mu \qquad (\varepsilon\ll1)
\end{equation}
yields the following modified algebra for the operators $\hat{x}^{\prime\mu}$:
\begin{equation}
 [\hat{x}^{\prime\mu},\hat{x}^{\prime\nu}]
 =i(1+2\varepsilon)\sigma^{\mu\nu}+\mathcal{O}(\varepsilon^2).
\end{equation}
This means that the change in the deformation parameter induced by 
infinitesimal
dilatations is given by $\delta\sigma^{\mu\nu}=2\sigma^{\mu\nu}$.
Therefore one expects the following relation:
\begin{equation}
 \int dx\,\delta_D\phi{\delta S^{(0)}\over\delta\phi}+\delta\sigma^{\mu\nu}
 {\partial S^{(0)}\over\partial\sigma^{\mu\nu}}=0.
\end{equation}
This reproduces exactly the result (\ref{l75}), (\ref{l76}).

\section{Conclusion and Outlook}
In the previous sections we have shown that one is able to construct an 
energy-momentum tensor which allows to define a conserved four momentum if
$\sigma^{0i}=0$. We have also demonstrated that the Noether theorem for
translations exists already at the operator level in terms of the operators
$\phi(\hat{x})$. Using the Moyal-Weyl correspondence between operators 
$\phi(\hat{x})$ and fields $\phi(x)$ we have also derived the energy momentum
tensor in the presence of a Moyal deformed interaction. Our result confirms
the results of \cite{r1,r2}.

In the last section we have also considered the dilatation symmetry directly
in a deformed field theory. We found that the Ward-identity of dilatation
symmetry picks up a breaking proportional to the deformation parameter
$\sigma^{\mu\nu}$. All our considerations are classical, i.e. without inclusion
of radiative corrections. Our investigations may be the basis to study the 
trace anomaly at least at the one loop level. In a further work \cite{r11} we
will try to give an answer whether the well-known trace anomaly \cite{r10} is modified
in a Moyal-Weyl deformed scalar quantum field theory.


\begin{thebibliography}{99}
\bibitem{r1}
 T.~Krajewski,
 ``Noncommutative geometry and fundamental interactions. (In French)'',
 math-ph/9903047.
\bibitem{r2}
 A.~Micu,
 ``Noncommutative $\phi^4$ theory at two loops'',
 hep-th/0008057.
\bibitem{r3}
 N.~L.~Balazs and B.~K.~Jennings,
 ``Wigner's Function And Other Distribution Functions In Mock Phase Spaces'',
 Phys.\ Rept.\  {\bf 104} (1984) 347.
\bibitem{r4}
 T.~Filk,
 ``Divergencies in a field theory on quantum space'',
 Phys.\ Lett.\  {\bf B376} (1996) 53.
\bibitem{r5}
 J.~Ambjorn, Y.~M.~Makeenko, J.~Nishimura and R.~J.~Szabo,
 ``Lattice gauge fields and discrete noncommutative Yang-Mills theory'',
 JHEP {\bf 0005} (2000) 023
 [hep-th/0004147].
\bibitem{r6}
 J.~Madore, S.~Schraml, P.~Schupp and J.~Wess,
 ``Gauge theory on noncommutative spaces'',
 Eur.\ Phys.\ J.\  {\bf C16} (2000) 161
 [hep-th/0001203].
\bibitem{r7}
 L.~Alvarez-Gaume and S.~R.~Wadia,
 ``Gauge theory on a quantum phase space'',
 hep-th/0006219.
\bibitem{r8}
 S.~Coleman and R.~Jackiw,
 ``Why Dilatation Generators Do Not Generate Dilatations?'',
 Annals Phys.\  {\bf 67} (1971) 552.
\bibitem{rj}
 R.~Jackiw,
 ``Dilatation symmetry and light-cone commutators'',
 Extended version of lectures delivered at the University of Turin, June,~1971;
 and at the DESY Summer School, July, 1971.
\bibitem{r9}
 J.~Gomis and T.~Mehen,
 ``Space-time noncommutative field theories and unitarity'',
 hep-th/0005129.
\bibitem{r10}
 M.~Schweda, Lett. al Nuovo Cimento, Vol. 9, No. 15 (1973) 596-598.
\bibitem{r11} 
 in preparation.
\bibitem{r12}
 T.~Pengpan and X.~Xiong,
 ``A note on the non-commutative Wess-Zumino model'',
 hep-th/0009070.
\end{thebibliography}
\end{document}